\documentclass[a4paper]{JHEP3-1-4}

\usepackage[english]{babel}
\usepackage{amsmath,amssymb}
\usepackage{graphicx}
\usepackage{ifthen,array}
\usepackage[comma,square,sort&compress]{natbib}
\usepackage{amsfonts}
\usepackage{bbm}

\allowdisplaybreaks

\newcommand{\be}{\begin{equation}}
\newcommand{\ee}{\end{equation}}
\newcommand{\bea}{\begin{eqnarray}}
\newcommand{\eea}{\end{eqnarray}}

\newcommand{\lp}{\left(}
\newcommand{\rp}{\right)}

\DeclareMathOperator{\imag}{Im}

\title{Stability and leptogenesis in the left-right symmetric seesaw mechanism}

\author{Evgeny Akhmedov\thanks{On leave from the National Research
Centre Kurchatov Institute, Moscow, Russia}, Mattias Blennow, Tomas
H\"allgren, Thomas Konstandin, and Tommy Ohlsson\\ Department of
Theoretical Physics, School of Engineering Sciences\\ Royal Institute
of Technology (KTH)\\ AlbaNova University Center \\ Roslagstullsbacken
21, 106 91 Stockholm, Sweden\\ E-mail:
\email{akhmedov@ictp.trieste.it}, \email{emb@kth.se},
\email{tomashal@kth.se}, \email{konstand@kth.se}, 
\email{tommy@theophys.kth.se}}

\abstract{We analyze the left-right symmetric type I+II seesaw
mechanism, where an eight-fold degeneracy among the mass matrices of
heavy right-handed neutrinos $M_R$ is known to exist. Using the
stability property of the solutions and their ability to lead to
successful baryogenesis via leptogenesis as additional criteria, we
discriminate among these eight solutions and partially lift their
eight-fold degeneracy. In particular, we find that viable leptogenesis
is generically possible for four out of the eight solutions.}


\preprint{}

\begin{document}

\section{Introduction}

In recent years, it has become an established fact that neutrinos,
though relatively light, are massive. Since the first experimental
evidence of neutrino oscillations until today an enormous progress has
been made in determining the low-energy properties of neutrinos, such
as mass squared differences and mixing. The existence of neutrino
masses poses some fundamental theoretical challenges, such as
understanding why the neutrino mass is so much smaller than the masses
of the other fermions. An elegant and attractive solution to this
problem is given by the seesaw mechanism~\cite{Minkowski:1977sc,
Gell-Mann:1980vs,Yanagida:1979as,Glashow:1979nm,
Mohapatra:1979ia,Magg:1980ut,Lazarides:1980nt,Schechter:1980gr,Mohapatra:1980yp},
which explains the smallness of the neutrino mass through the
existence of very heavy particles (usually right-handed Majorana
neutrinos or Higgs triplets), the mass scale of which could be related
to that of Grand Unification. In addition, the seesaw mechanism
provides a natural explanation of the baryon asymmetry of the Universe
through the baryogenesis via leptogenesis
mechanism~\cite{Fukugita:1986hr} (for recent reviews, see
refs.~\cite{Buchmuller:2004tu,Buchmuller:2005eh,Strumia:2006qk}).
However, the large mass scale of the seesaw particles jeopardizes the
hopes of testing this mechanism in the laboratory and hence reduces
its predictivity.

In the present work, we consider the seesaw mechanism in a class of
left-right symmetric models in which the intermediate states with both
right-handed neutrinos (type I) and heavy triplet scalars (type II)
contributions to the light neutrino mass matrix $m_\nu $ are naturally
present. We focus on a special case with a discrete left-right
symmetry, in which type I and type II seesaw contributions contain the
same triplet Yukawa coupling $f$. This case has much fewer parameters
than the most general one and is therefore more predictive. After
integrating out the heavy particles, the light neutrino mass matrix is
given by
\be
m_\nu = f \, v_L - \frac{v^2}{v_R} \, y \, f^{-1} y^T\,, 
\label{eq:ss1}
\ee
where $f$ is the triplet Majorana-type Yukawa coupling, $y$ is the
Dirac-type Yukawa coupling of neutrinos and $v$, $v_L$, and $v_R$ are
vacuum expectation values (VEVs). The first term in eq.~(\ref{eq:ss1})
is the type II contribution, while the second term is the type I
contribution from the original seesaw scenario.  In the case when $y$
is a complex symmetric matrix, it was shown in
ref.~\cite{Akhmedov:2005np} that if the light neutrino mass matrix
$m_\nu$, the VEVs, and the Dirac-type Yukawa coupling matrix $y$ are
known, the seesaw formula (\ref{eq:ss1}) can be inverted analytically
to find the triplet Yukawa coupling matrix $f$. Since the seesaw
equation is non-linear in $f$, one can expect multiple solutions, and
indeed an eight-fold of allowed solutions is found
\cite{Akhmedov:2005np}.  As the mass matrix of heavy right-handed
Majorana neutrinos is given by $M_R=f v_R$, this also implies an
eight-fold ambiguity for this mass matrix.  For given Dirac-type
Yukawa coupling matrix $y$ and VEVs, all eight solutions for $f$
result in exactly the same mass matrix of light neutrinos $m_\nu$, and
thus, the seesaw relation by itself does not allow one to select the
true solution among the possible ones. One therefore has to invoke
some additional information and/or selection criteria. The present
work is an attempt in this direction.

One possibility to discriminate among the eight allowed solutions
for $f$ is to introduce a notion of naturalness. For example, for
certain ranges of the VEVs and certain solutions, a very special
triplet coupling matrix $f$ might be needed, in the sense that
marginally different $f$ would lead to significantly
different low-energy phenomenology. We consider such a situation
unnatural; the degree of tuning that is required in the right-handed
sector to obtain the observed neutrino phenomenology will be
quantified and the corresponding selection criterion for $f$ discussed
in section~\ref{sec_stab}.

Another possibility to discriminate among the allowed solutions is
to constrain them by the phenomenology of the right-handed
neutrinos. Since the right-handed sector of the theory is not directly
accessible to laboratory experiments, cosmological benchmarks turn out
to be the most promising tool.  Namely, we will classify the solutions
according to their ability to lead to successful baryogenesis via
leptogenesis. This will be discussed in section~\ref{sec_lep}, before
we draw our conclusions in section~\ref{sec_summary}.

Recently, leptogenesis in a class of models with the left-right symmetric
seesaw mechanism has been considered in a similar framework in
ref.~\cite{Hosteins:2006ja}. We compare our results with those in
ref.~\cite{Hosteins:2006ja} in sections~\ref{sec_lep}
and~\ref{sec_summary}. 

\section{The model and the inversion formula}
\label{sec_model}

In this section, we introduce our framework and set up the notation.
In the basis where the mass matrix of charged leptons is diagonal, the
light neutrino mass matrix can be written as
\be
m_\nu = (P_{l} \, U_{\rm PMNS} \, P_\nu)^* \, m_{\nu}^{\rm diag}\,
(P_{l} \, U_{\rm PMNS} \, P_\nu)^\dag\,,  
\ee
where $m_{\nu}^{\rm diag}={\rm diag}(m_1,\,m_2,\,m_3)$ is the diagonal
matrix of neutrino masses, $U_{\rm PMNS}$ is the leptonic mixing
matrix which depends on three mixing angles and a Dirac-type
CP-violating phase, and $P_l$ and $P_\nu$ are diagonal matrices of
phase factors, which in general contain five independent complex
phases.

The neutrino masses $m_1$, $m_2$, and $m_3$ can be expressed through
the lightest neutrino mass $m_0$ and the two mass squared differences
$\Delta m_{21}^2$ and $\Delta m_{31}^2$. In our numerical calculations,
we will use the current best-fit values of the parameters defining the
neutrino mass
matrix~\cite{Maltoni:2004ei,Strumia:2005tc,Fogli:2006qg}:
\be
\label{pmns_1}
\Delta m^2_{21} \simeq 7.9 \times 10^{-5}~\textrm{eV}^2\,, \quad
\Delta m^2_{31} \simeq \pm 2.6 \times 10^{-3}~\textrm{eV}^2\,,
\ee
\be
\quad\theta_{12} \simeq 33.2^\circ\,, \quad 
\theta_{23} \simeq 45^\circ\,.
\ee
For the mixing angle $\theta_{13}$, only the upper limit $\theta_{13}
\lesssim 11.5^\circ$ exists. Unless explicitly stated otherwise, 
we will use the value $\theta_{13}=0$ in our analysis.

We will be assuming that the Dirac-type Yukawa coupling matrix of
neutrinos $y$ coincides with that of the up-type quarks $y_u$. This is
a natural choice in the light of quark-lepton symmetry and grand
unified theories (GUTs)
\cite{Pati:1973uk,Georgi:1974my,Fritzsch:1974nn}. 
Actually, this relation is unlikely to hold exactly, since, in the GUT
framework, it would also imply that the Yukawa couplings of the
down-type quarks and charged leptons coincide, $y_d=y_l$, in
contradiction with experiment. GUT models that modify this relation
usually also modify the relation between the up-type and neutrino
Yukawa matrices~\cite{Babu:1992ia, Anderson:1993fe}.  However, most of
the qualitative results in the present work depend only on the fact
that the eigenvalues of $y$ are hierarchical. Whenever a result relies
on the assumption $y=y_u$, we will comment explicitly on this
issue. Following ref.~\cite{Akhmedov:2005np}, we will also assume $y$
to be symmetric.  In this case, the two VEVs ($v_L$ and $v_R$), the
sign of $\Delta m_{31}^2$, and the mass scale of the light neutrinos
are the only free parameters (ignoring for the moment the CP-violating
phases, which will be discussed in section~\ref{sec_lep}).

Our choice of the Dirac-type Yukawa coupling matrix implies that it
can be written as
\be
y = P_{d} \, U_{\rm CKM}^T \,  P_{u} \,  y_{u}^{\rm diag} 
\, P_{u} \, U_{\rm CKM} \, P_{d}\,,
\label{y}
\ee
where the eigenvalues of $y_{u}^{\rm diag}$ are
\be
\label{ckm_1}
y_u^{\rm diag} = {\rm diag}(4.2 \times 10^{-6},\, 1.75 \times 10^{-3}, 
\,0.7)\,,
\ee
and we use the standard parameterization of the CKM matrix $U_{\rm CKM}$
\cite{Yao:2006px} with
\be
\label{ckm_2}
\theta^q_{12} \simeq 13.0^\circ, \quad 
\theta^q_{13} \simeq 0.2^\circ,\quad \theta^q_{23} \simeq 2.2^\circ, \quad 
\delta^q \simeq 1.05\,.
\ee
The values in eqs.~(\ref{ckm_1}) and (\ref{ckm_2}) are evaluated at
the GUT scale, following ref.~\cite{Hosteins:2006ja}.
The matrices $P_{u}$ and $P_{d}$ in eq.~(\ref{y}) are
diagonal matrices of phase factors.  The phases in the four matrices
$P_{l}$, $P_\nu$, $P_u$, and $P_{d}$ are partially redundant. For
example, by a redefinition of the fields, the three phases of $P_{l}$
can be moved into $P_{d}$, so that we are left with the two usual
Majorana phases and the Dirac phase in the low-energy sector, while
five additional Majorana phases and one Dirac phase reside in $y$ and
can only affect high-energy processes such as leptogenesis. Even
though these phases can marginally influence the stability of the
seesaw solutions, we set the high-energy phases to zero in the first
part of our work and consider them only in the part where leptogenesis
is discussed.

In order to invert the seesaw formula, it is useful to introduce the
following dimensionful quantities:
\be
g = v_L \, f\,, \qquad \mu = \frac{v_R}{v_L \, v^2}\,,
\ee
with the VEV $v \simeq 174$~GeV, so that eq.~(\ref{eq:ss1}) turns into
\be
m_\nu = g - \frac{1}{\mu} \, y \, g^{-1} y^T. 
\label{eq:ss2}
\ee
This convention has the advantage that the matrix $g$ will only depend
on $\mu$ and not on the two VEVs, $v_L$ and $v_R$, separately. It will
turn out that the baryon asymmetry produced via leptogenesis depends
only on this combination of VEVs, so that, besides the CP-violating
phases, we are left with two parameters only, the quantity $\mu$ and
the lightest neutrino mass $m_0$. The hierarchy of the light neutrino
masses can be considered as an additional discrete parameter.

In the following, we give a short description of the seesaw inversion
formula from refs.~\cite{Akhmedov:2005np,Akhmedov:2006de} in the case
of three lepton generations and when $y$ is a complex symmetric
matrix. In the basis where $y$ is diagonal, the seesaw equation for
$g$ reduces to the following system of six coupled non-linear
equations for its matrix elements $g_{ij}$:
\begin{equation}\label{eq:coupledeqs}
\mu G [ g_{ij}-(m_{\nu})_{ij} ]=y_{i}y_{j}G_{ij}\,.
\end{equation} 
Here we use the notation  
\be
G \equiv {\det}\,g, \qquad
G_{ij}=\frac{1}{2} \sum_{k,l,m,n=1}^3\epsilon_{ikl}
\epsilon_{jmn}g_{km}g_{ln}\,.
\ee
It was found in ref.~\cite{Akhmedov:2005np} that in the case when $y$
is symmetric, for every solution $g$ there exists another solution
$\tilde{g}$ which is related to $g$ by the duality transformation
$\tilde{g}=m_{\nu}-g$. For $\tilde{g}$, eq.~(\ref{eq:coupledeqs}) reads
\begin{equation}
\mu\tilde{G}[\tilde{g}_{ij}-(m_{\nu})_{ij}]=-\mu\tilde{G}g_{ij}=y_{i}y_{j}
\tilde{G}_{ij}
\end{equation}
with $\tilde{G}\equiv {\det}\,\tilde{g}$. The system of equations in
eq.~(\ref{eq:coupledeqs}) can now be solved by making use of the
following procedure. First, we introduce the rescaled matrices
$g'=g/\lambda^{1/3}$, $m_{\nu}'=m_{\nu}/\lambda^{1/3}$, and
$y'=y/\lambda^{1/3}$, where $\lambda$ is to be determined from the
equation $G'(\lambda)\equiv {\det}\, g'(\lambda)=1$. Then, using
the equation for the dual quantities $\tilde g'$, one can linearize the
system of equations for $g'_{ij}$. Next, this system can be solved and
one obtains the following solution for $g$:
\begin{equation}\label{eq:g_{ij}}
g_{ij}=\frac{\lambda^{2}[(\lambda^{2}-Y^{2})^{2}-Y^{2}\lambda \,{\rm
det}\,m_{\nu} +
Y^{4}S](m_{\nu})_{ij}+\lambda(\lambda^{4}-Y^{4})A_{ij}-Y^{2}\lambda^{2}
(\lambda^{2}+Y^{2})S_{ij}}{(\lambda^2-Y^2)^3-Y^2\lambda^2(\lambda^{2}-Y^{2})
S-2Y^{2}\lambda^{3}\,{\det}\, m_{\nu}}\,,
\end{equation}
where 
\begin{equation}
Y^{2}\equiv\frac{(y_{1}y_{2}y_{3})^{2}}{\mu^{3}},\quad S\equiv \mu
\sum_{k,l=1}^{3}\left[\frac{(m_{\nu})_{kl}^{2}}{y_{k}y_{l}}\right],
\quad A_{ij}\equiv\frac{y_{i}y_{j}M_{ij}}{\mu},\quad
\ee
\be
S_{ij}\equiv \mu \sum_{k,l=1}^{3}\left[(m_{\nu})_{ik}(m_{\nu})_{jl}
\frac{(m_{\nu})_{kl}}{y_{k}y_{l}}\right]
\end{equation}
with $M_{ij}=\frac{1}{2}\epsilon_{ikl}\epsilon_{jmn}(m_{\nu})_{km}
(m_{\nu})_{ln}$. In terms of the original (non-rescaled) quantities,
one has $G(\lambda)\equiv{\det}\,g(\lambda)=\lambda$, which yields
an eighth order equation for $\lambda$. Using the duality property,
one can reduce it to a pair of fourth order equations. Substituting the
solutions for $\lambda$ into eq.~(\ref{eq:g_{ij}}) gives eight
solutions for $g_{ij}$. In general, for $n$ lepton generations the
number of solutions is $2^n$ \cite{Akhmedov:2005np}.

The matrix structure of the solutions of the seesaw equation was
studied in some detail in ref.~\cite{Akhmedov:2006de}. In the present
work, we will rather focus on the eigenvalues of the matrices $g$,
the corresponding mixing parameters, stability properties of the
solutions, and the implications for leptogenesis.

\section{Stability analysis\label{sec_stab}}

Since the neutrino Dirac-type Yukawa coupling matrix in our framework
is given by the up-type quark mass matrix, the inversion formula of
the previous section can be used to determine the eight allowed
structures of the triplet coupling matrix $f=g/v_L$ for given
low-energy neutrino mass matrix $m_\nu$ and the parameters $v_L$,
$v_R$, and $m_0$. Our stability analysis is based on the assumption
that the Dirac-type coupling matrix $y$ and the Majorana-type coupling
matrix $f$ are {\it a priori} independent (for a discussion of the
situations when this is not the case, see section 5 of
ref.~\cite{Akhmedov:2006de}). We pose the question of whether the
resulting low-energy phenomenology is stable under small changes in
$f$. Since the inversion formula in general yields eight valid
solutions, the mass matrix $m_\nu$ and the corresponding Majorana
coupling matrix $f$ are in a 1-to-8 correspondence.  It is still a
reasonable question to ask if for the measured $m_\nu$ some of the
predicted $f$ have to be very special, so that a fine-tuning is
required and a small modification of their elements may lead to a
large change in $(m_\nu)_{ij}$.

The measure we use to quantify the stability property of the solutions
is the following:
\be
\label{Q_measure}
Q = \left| \frac{\det{f}}{\det{m_\nu}} \right|^{1/n} \sqrt{\sum_{k,l=1}^{2N} 
\left( \frac{\partial {m_l}}{\partial f_k} \right)^2}\,.
\ee
The real coefficients $f_k$ and $m_l$ determine the matrices $f$
and $m_\nu$ according to
\bea
f &=& \sum_k ( f_k + i f_{k+N}) T_k, \\
m_\nu &=& \sum_k ( m_k + i m_{k+N}) T_k, 
\eea
where $T_k$, $k\in[1,N]$ with $N=n(n+1)/2$, form a basis of complex 
symmetric $n\times n$ matrices.
For this basis, we choose the normalization
\be
\label{T_normalize}
{\rm tr \,} ( T^\dagger_l \, T_k ) = \delta_{lk}\,.  
\ee
The resulting stability measure $Q$ does not depend on the chosen
basis.  This can be easily seen in the following way. Consider another
basis $T_k^\prime$ satisfying eq.~(\ref{T_normalize}). The two bases
are then connected via a unitary transformation $T_k^\prime=\sum_l
U_{kl}
\, T_l$.  The coefficients in the old and new bases are determined as
\bea
f_k = {\rm \, Re \,} \left[ {\rm \, tr \,} (T^\dagger_k\, f) \right], &&
f_{k+N} = {\rm \, Im \,} \left[ {\rm \, tr \,} (T^\dagger_k\, f) \right], \\ 
f^\prime_k = {\rm \, Re \,} \left[ {\rm \, tr \,} (T^{\prime\dagger}_k\, f) 
\right], && f^\prime_{k+N} = {\rm \, Im \,} \left[ {\rm \, tr \,} 
(T^{\prime\dagger}_k\, f) \right], 
\eea
and hence, are related by an orthogonal transformation
\be
f_a^\prime = \sum_b O_{ab} \, f_b, \quad
a,b \in[1,2N], \quad
O = 
\begin{pmatrix}
{\rm \,Re\,} U & {\rm \,Im\,} U \\
-{\rm \,Im\,} U & {\rm \,Re\,} U \\
\end{pmatrix},
\ee
which leaves the measure in eq.~(\ref{Q_measure}) invariant~\footnote{
Note that the stability issue was also discussed in
ref.~\cite{Hosteins:2006ja} where a different stability criterion,
constraining only the element $f_{33}$, was introduced. }.

Many interesting properties of the seesaw inversion formula appear
already in the one-flavor case. The solutions $g$ are then given by
\be
g = \frac{m_\nu}{2} \pm \sqrt{\frac{m^2_\nu}{4} + \frac{y^2}{\mu}}
\ee
and our stability measure simplifies to
\be
Q = f \, \frac{d}{d f} \log{|m_\nu|} = 
g \, \frac{d}{d g} \log{|m_\nu|} 
= \sqrt{1 + \frac{4\, y^2}{\mu m_\nu^2}}\;.~~~
\label{Q1}
\ee
In the following, we will discuss the qualitative behavior of the
solutions $f$ in various regions of the parameter space and its
implications for the stability of these solutions. 

\FIGURE[t]{
\includegraphics[width=0.7\textwidth,clip]{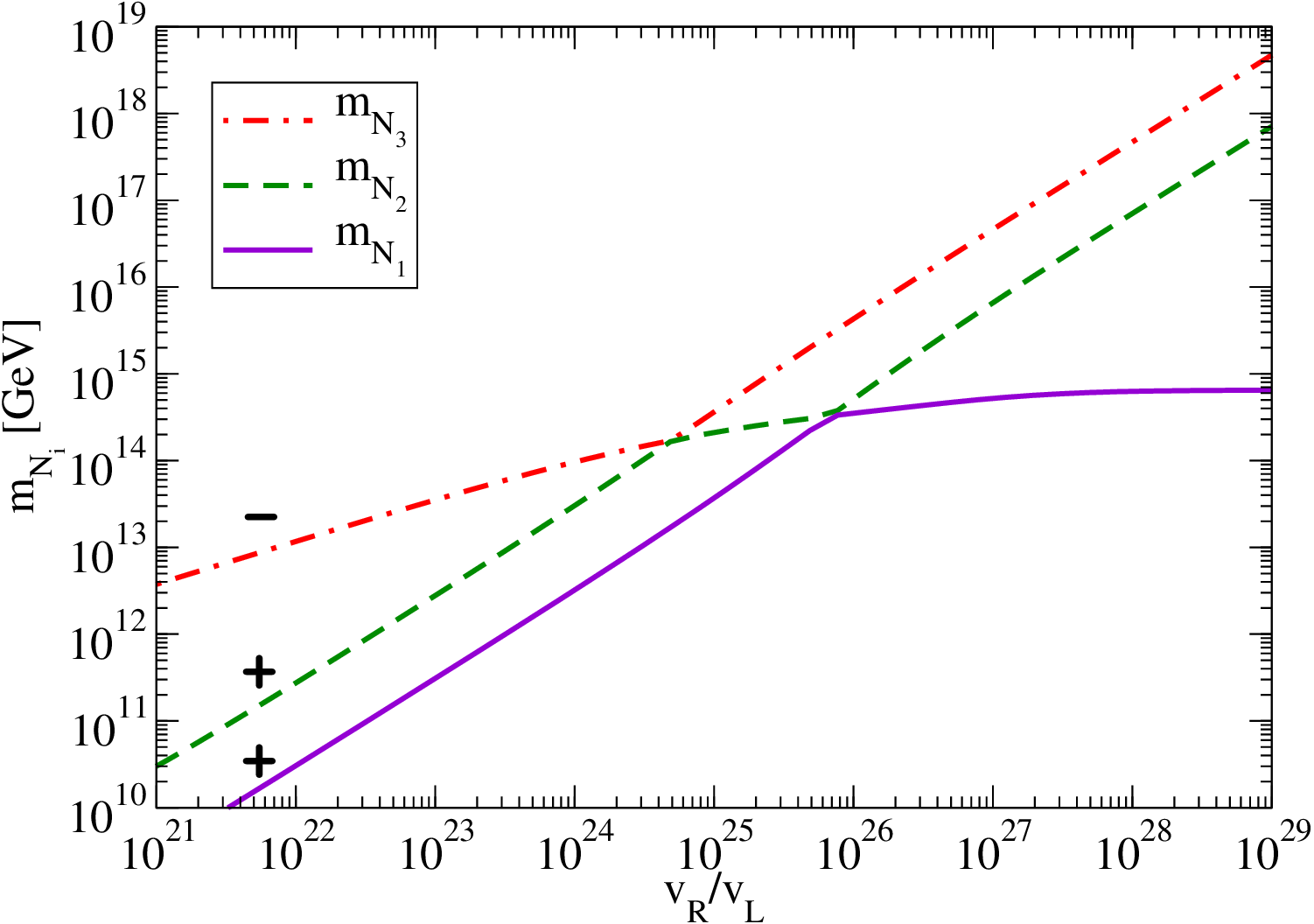}
\caption{An example of our labeling convention for the solution
'$-++$'.}
\label{ex_conv}
}
\FIGURE[t]{
\includegraphics[width=0.9 \textwidth,clip]{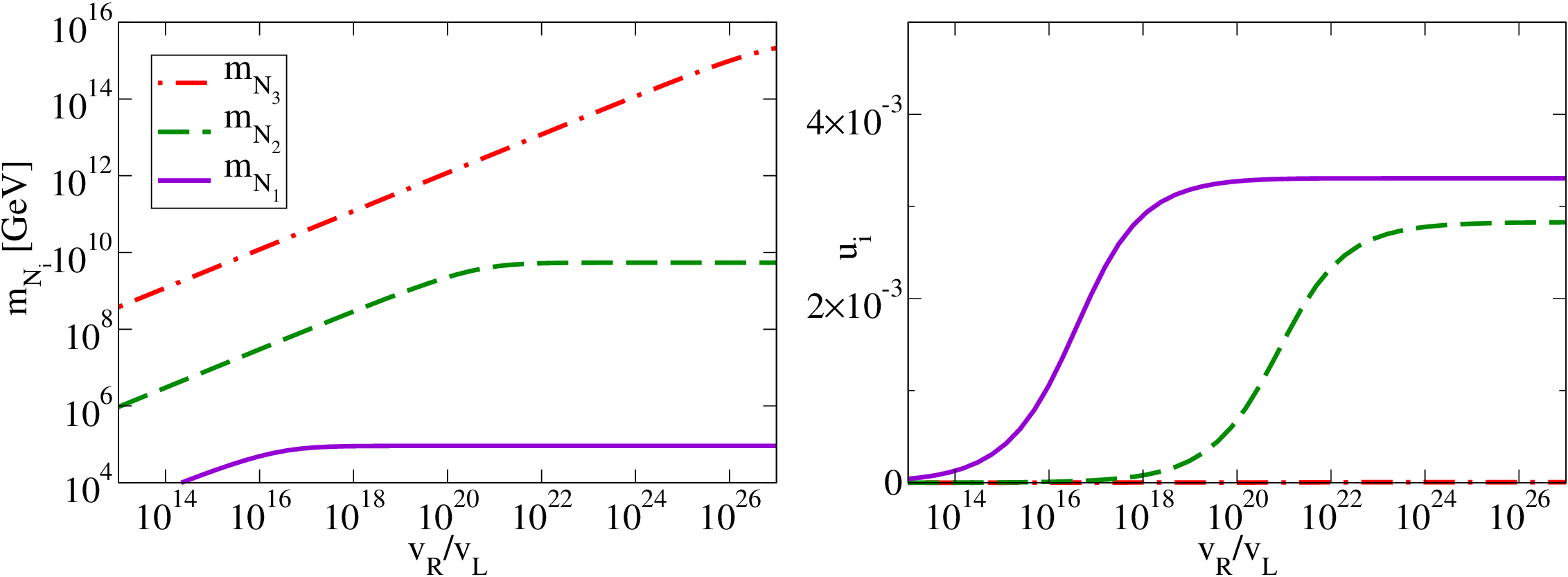}
\caption{The right-handed neutrino masses $m_{N_i}$ and mixing
parameters $u_i$ as functions of $v_R/v_L$ for the solution
'$---$'. Normal mass hierarchy, $m_0=0.001$~eV.}
\label{ex_1}
}
\FIGURE[t]{
\includegraphics[width=0.9 \textwidth,clip]{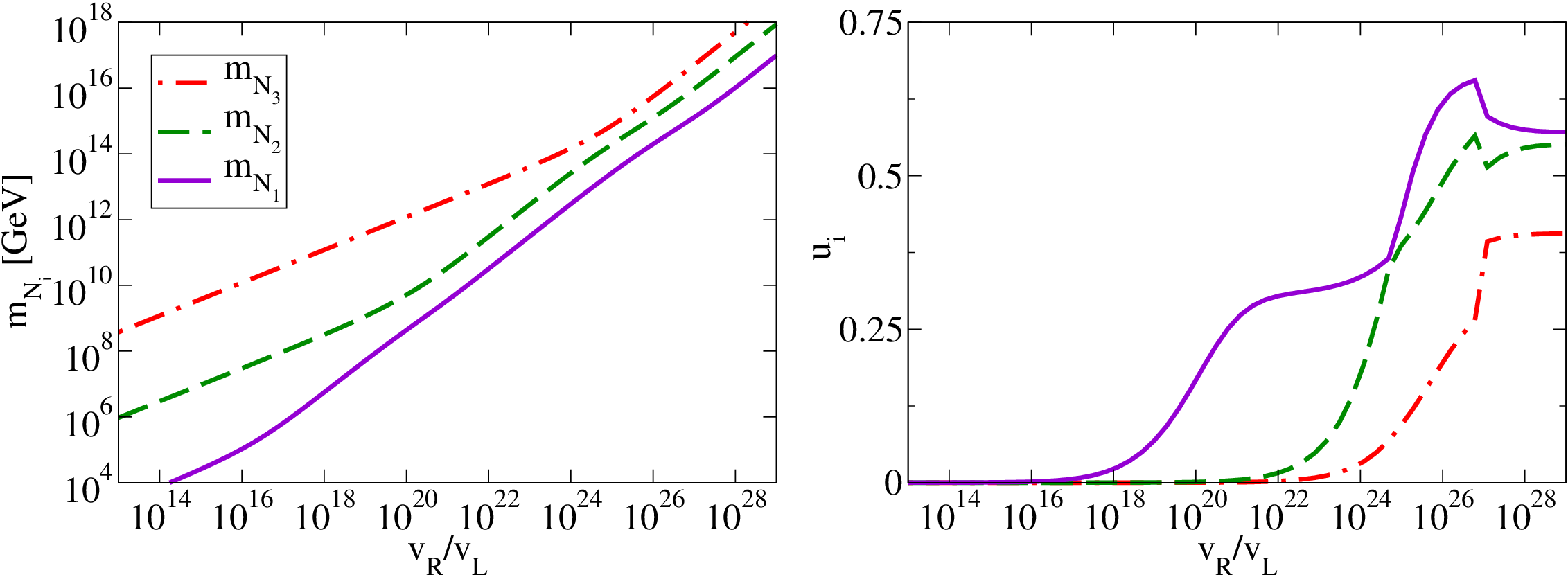}
\caption{Same as in fig.~\ref{ex_1}, but for the solution '$+++$'.}
\label{ex_2}
}
\FIGURE[t]{
\includegraphics[width=0.9 \textwidth,clip]{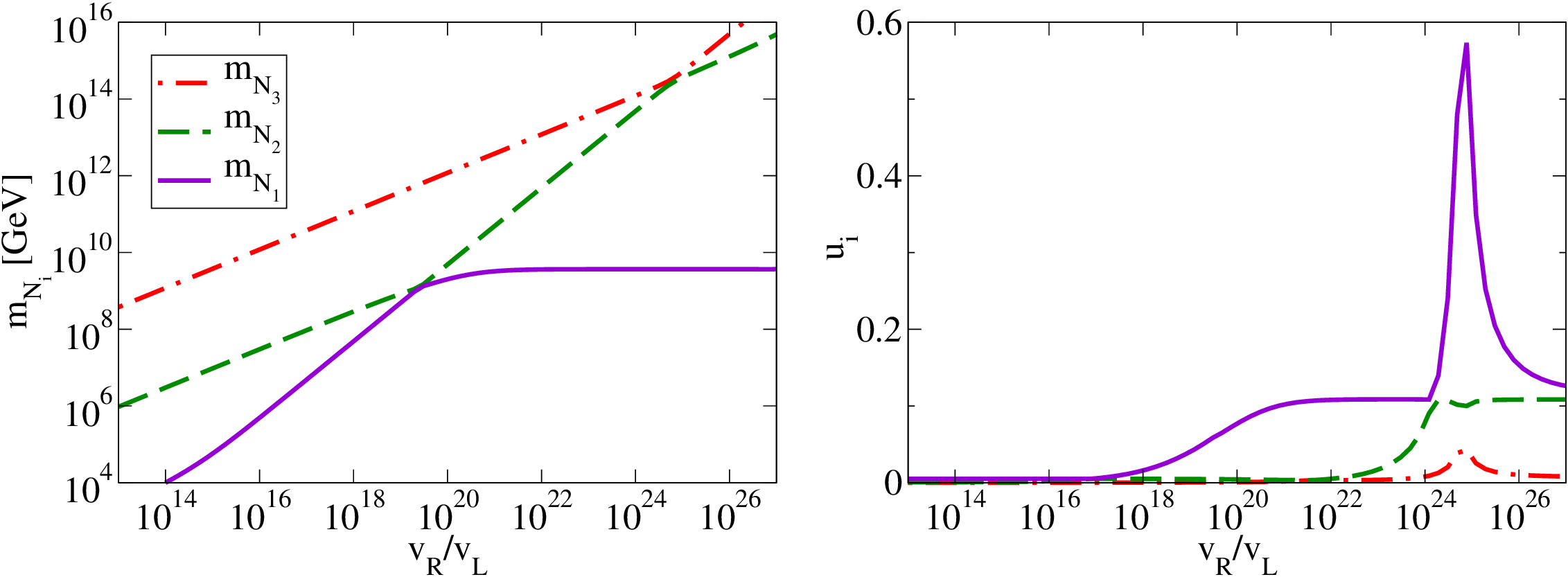}
\caption{The right-handed neutrino masses $m_{N_i}$ and mixing
parameters $u_i$ as functions of $v_R/v_L$ for the solution
'$+-+$'. Inverted mass hierarchy, $m_0=0.001$~eV.}
\label{ex_3}
}
\FIGURE[t]{
\includegraphics[width=0.9 \textwidth,clip]{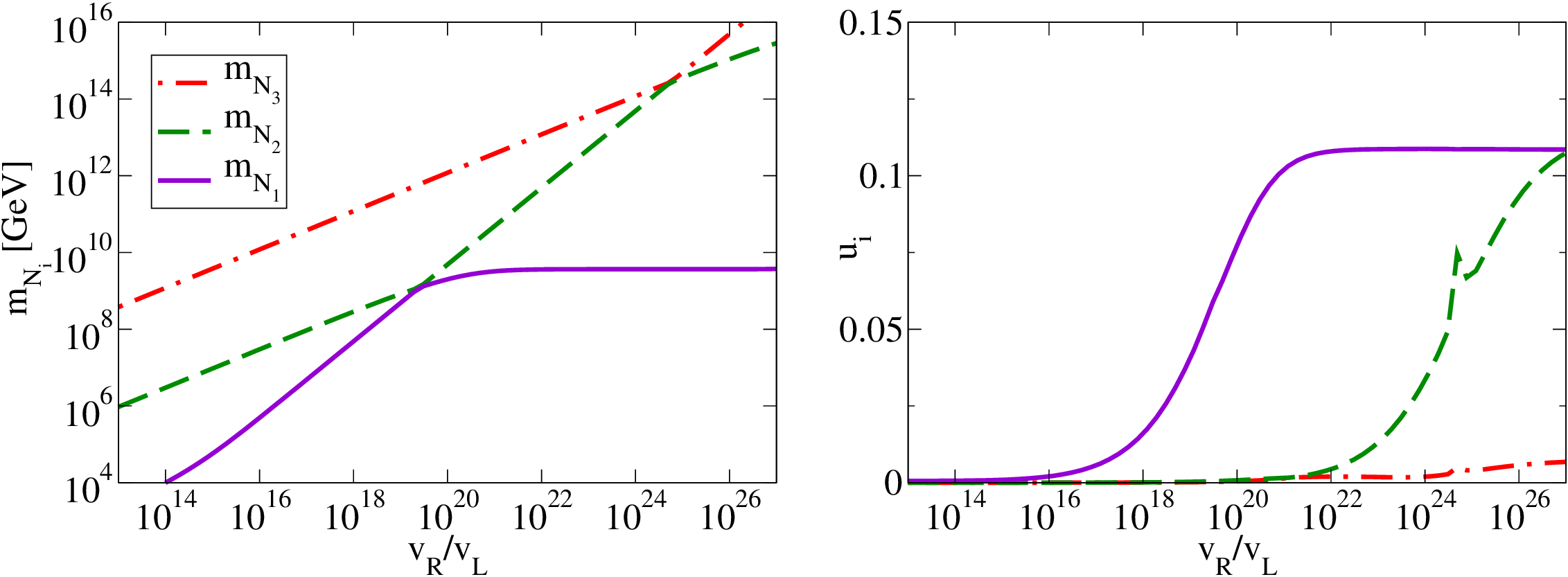}
\caption{Same as in fig.~\ref{ex_3}, but for the solution '$--+$'.}
\label{ex_4}
}

\subsection[Large $\mu$ regime]{Large $\boldsymbol{\mu}$ regime}

In the regime of large $\mu$, 
\be
\mu \gg \frac{4y^2}{m_\nu^2}\,, 
\label{large_mu}
\ee
the two solutions in the one-flavor case are given by
\be
g \to - \frac{y^2}{\mu m_\nu} \quad \textrm{and} \quad g \to m_\nu\,. 
\ee
In this regime, the solutions are purely type I or type II dominated.
In the three-flavor case, the eight solutions follow from the eight
corresponding choices for the eigenvalues and we will label these
solutions according to their limiting behavior at large $\mu$ as '$-$'
or '$+$' in the case of type I or type II dominance (starting with
the largest eigenvalue in the small $\mu$ regime). This notation
agrees with the one used in ref.~\cite{Hosteins:2006ja}.
Our convention is illustrated in fig.~\ref{ex_conv} using the solution
'$-++$' as an example.

From eq.~(\ref{Q1}) one can observe that in the large $\mu$ regime
of the one-flavor case, both solutions for $g$ are characterized by
the stability measure $Q \simeq 1$, which is a very stable
situation. Note that for the three-flavor case, no fine-tuning
corresponds to $Q \simeq 10$. However, for several flavors and
hierarchical $y$, there is in general an instability related to mixing
that will be discussed in the next subsection.

\subsection{Hierarchy induced large mixing}

For simplicity, we start with a discussion of the two-flavor case in
the pure type I seesaw framework. By hierarchy induced large mixing we
mean the following: Suppose that $y$ has a hierarchical structure
\be
y \sim \begin{pmatrix}
\epsilon & 0 \\
0 & 1 \\
\end{pmatrix},
\ee
while, in contrast to this, the low-energy neutrino mass has a rather
mild or even no hierarchy. Then, the corresponding matrix $g$ is
necessarily characterized by the hierarchy that is the squared
hierarchy of $y$. Indeed, introducing a unitary matrix $U(\theta)$
that diagonalizes $g$, one finds
\be
g = -\frac1\mu {y \, m_\nu^{-1}\, y}=U^\dagger(\theta)\,\hat g \, U^*(\theta)
\ee
with the diagonal matrix 
\be
\hat g \sim 
\begin{pmatrix}
\epsilon^2 & 0 \\
0 & 1 \\
\end{pmatrix},
\ee
and, in addition, mixing has to be small, i.e.~$\theta \sim \epsilon$. This
was already observed in
refs.~\cite{Smirnov:1993af,Tanimoto:1995uw,Altarelli:1999dg} and
suggested as a possible mechanism for generating large mixing angles
in the light neutrino mass matrix out of small mixing angles in the
right-handed and Dirac sectors. However, in our context, this is not a
desirable situation, since it would require a fine-tuning between the
Dirac and Majorana Yukawa couplings, i.e.~between the sectors that we
have assumed to be unrelated.  In terms of stability, this would lead
to large values of $Q$. In addition, the large hierarchy among the
elements of the Dirac-type Yukawa coupling matrix $y$ would induce a
huge hierarchy among the elements of $g$, leading in general to an
extremely small mixing in the right-handed neutrino sector, which may
preclude successful leptogenesis.

The above consideration was based on the type I seesaw formula, and
hence, is not fully applicable to our framework. Still, it applies to
the solutions dominated by type I seesaw. Figure~\ref{ex_1} shows the
one out of the eight solutions that is fully dominated by the type I
term in the large $\mu$ regime and is labeled as '$---$'. As a measure
of mixing, we consider the parameters $u_i$ which are related to the
off-diagonal elements of the unitary matrix $U$ diagonalizing $g$ as
follows~\footnote{Recall that we work in the basis where the matrix
$y$ is diagonal.}:
\be
u_1^2=\frac12 (|U_{12}|^2 + |U_{21}|^2)\,, \quad 
u_2^2=\frac12 (|U_{13}|^2 + |U_{31}|^2)\,, \quad
u_3^2=\frac12 (|U_{23}|^2 + |U_{32}|^2)\,.
\ee
These parameters, along with the masses of right-handed neutrinos, are
plotted for several solutions in figs.~\ref{ex_1}-\ref{ex_4}.

For the solution '$---$', mixing is small in the large $\mu$ regime,
as can be seen from fig.~\ref{ex_1}. For the other seven solutions,
this does not hold in general, as can be seen e.g.~in fig.~\ref{ex_2}.
However, even in the general case, one feature seems to be universal:
If the matrix elements of $g$ exhibit a strong hierarchy, then the
mixing in the right-handed sector is suppressed, which leads to the
necessity of fine-tuning between the Dirac and Majorana sectors and
related instabilities. This also explains why the two solutions
'$++-$' and '$+--$' are very unstable with almost equal stability
measure $Q$. The strong hierarchy between the largest and smallest
right-handed masses leads to large instabilities, while the behavior
of the third mass is rather irrelevant.

\subsection[Small $\mu$ regime]{Small $\boldsymbol{\mu}$ regime}

When $\mu$ is small in the sense that 
\be
\mu \ll \frac{4y^2}{m_\nu^2}\,, 
\label{small_mu}
\ee
in the one-flavor case, one finds the following limiting behavior for $g$:
\be
g \to \pm \frac{y}{\sqrt{\mu}} + \frac{m_\nu}{2} + \mathcal{O}(\sqrt{\mu}), 
\quad \mu \to 0. \label{sol_small_mu}
\ee
For the stability measure, eq.~(\ref{Q1}) gives 
\be
Q = \left|\frac{g}{m_\nu} \frac{d m_\nu}{d g} \, \right| 
\to \frac{2y}{\sqrt{\mu} \, m_\nu} \to \infty
\ee
in this limit, and therefore a very unstable situation. This had to be
expected, since there is an almost exact cancellation between the type
I and type II contributions to $m_\nu$ in the seesaw formula in this
regime. In the multi-flavor case, there is an additional instability
in the small $\mu$ limit which stems from the fact that mixing in $g$
is suppressed by the hierarchy in $y$. This can be illustrated by the
two-flavor case, in which the four solutions are of the form
\be
g = \frac{1}{\sqrt{\mu}} y^{1/2} P y^{1/2}
\ee
with $P$ of the form
\be
P ~\propto ~\pm \mathbbm{1} + {\cal O}(\sqrt{\mu}) 
\quad {\rm or } \quad
P ~\propto ~\pm 
\begin{pmatrix}
\cos \alpha & \sin \alpha \\
\sin \alpha & -\cos \alpha \\
\end{pmatrix}
+ {\cal O}(\sqrt{\mu}). 
\ee
For the first pair of solutions, mixing vanishes in the limit $\mu \to
0$, while for the second pair, mixing in $g$ is suppressed by the
hierarchy in $y$.  A similar argument applies to the three-flavor case
and can be observed in our numerical results.  For example, this
behavior can be seen in figs.~\ref{ex_1} and \ref{ex_2} which display
two out of the eight solutions for the normal mass hierarchy and
$m_0=0.001$~eV.

\subsection{Numerical results}

Figures~\ref{stab_1} and~\ref{stab_2} show the stability measure $Q$
for small and large $m_0$ and normal/inverted mass hierarchy. For
small $m_0$, the transition from the large $\mu$ to the small $\mu$
regime appears for larger values of $\mu$, in accordance with
eqs.~(\ref{large_mu}) and (\ref{small_mu}). In all four scenarios, the
solutions are unstable in the regime of small $\mu$, which is due to
the cancellation between type I and type II contributions to the mass
matrix of light neutrinos. In addition, the solutions where the
smallest eigenvalue is dominated by type I seesaw in the large $\mu$
regime ('$\pm\pm-$'), are unstable for large $\mu$ as well, since the
lightest right-handed mass stays below $10^6$~GeV in this limit and
this generally leads to a large spread in the eigenvalues and to
instabilities, as explained in the previous sections. Examples of the
eigenvalues in these cases are given in fig.~\ref{ex_1}. Analogously,
the stability measure of the solutions '$\pm-+$' increases for
$v_R/v_L \gtrsim 10^{20}$, since the smallest right-handed neutrino
mass approaches its asymptotic value of about $10^9$~GeV, as can be
seen in figs.~\ref{ex_3} and \ref{ex_4}. A similar effect
appears for the solution '$-++$' at values $v_R/v_L \gtrsim 10^{24}$.
The purely type II dominated solution ('$+++$') is the most stable one
in almost all the cases.
\FIGURE[t]{
\includegraphics[width=0.49\textwidth,clip]{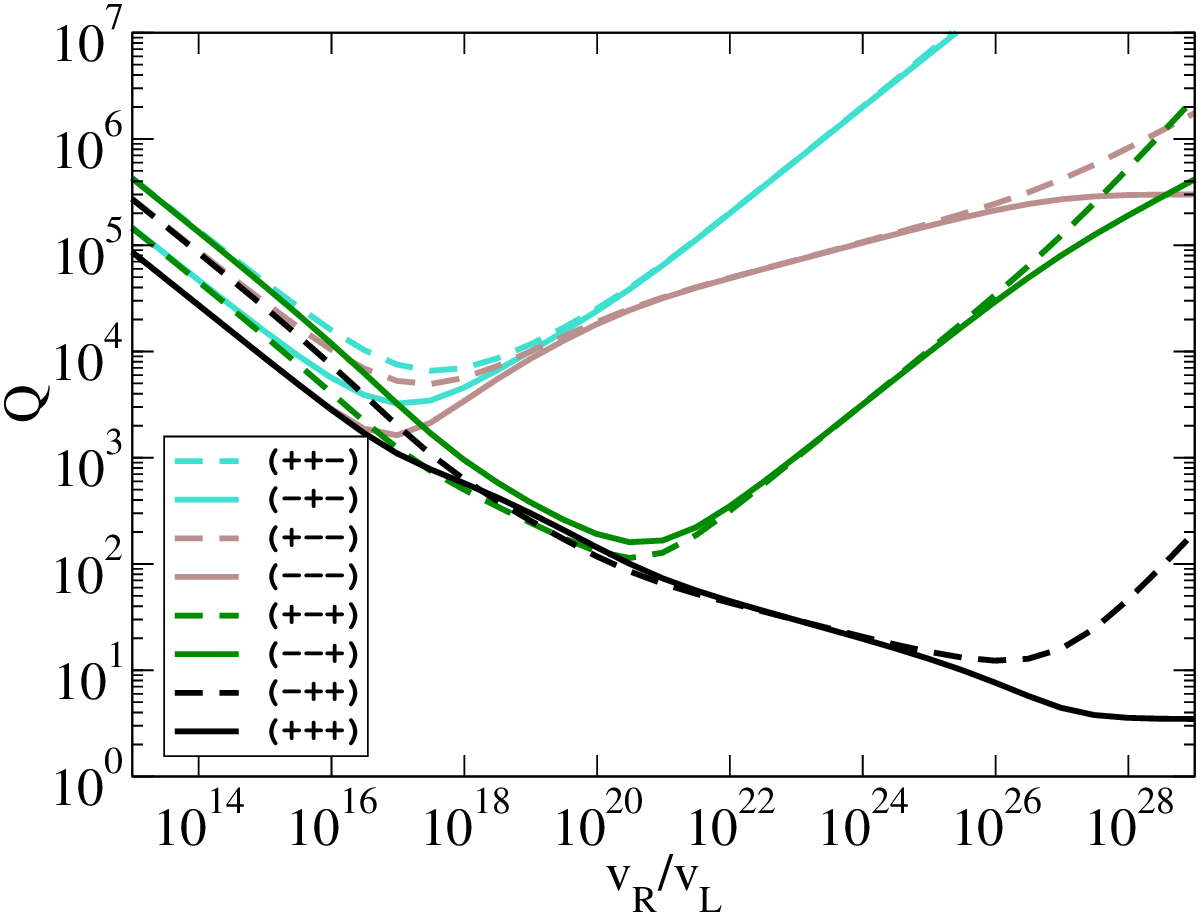}
\includegraphics[width=0.49\textwidth,clip]{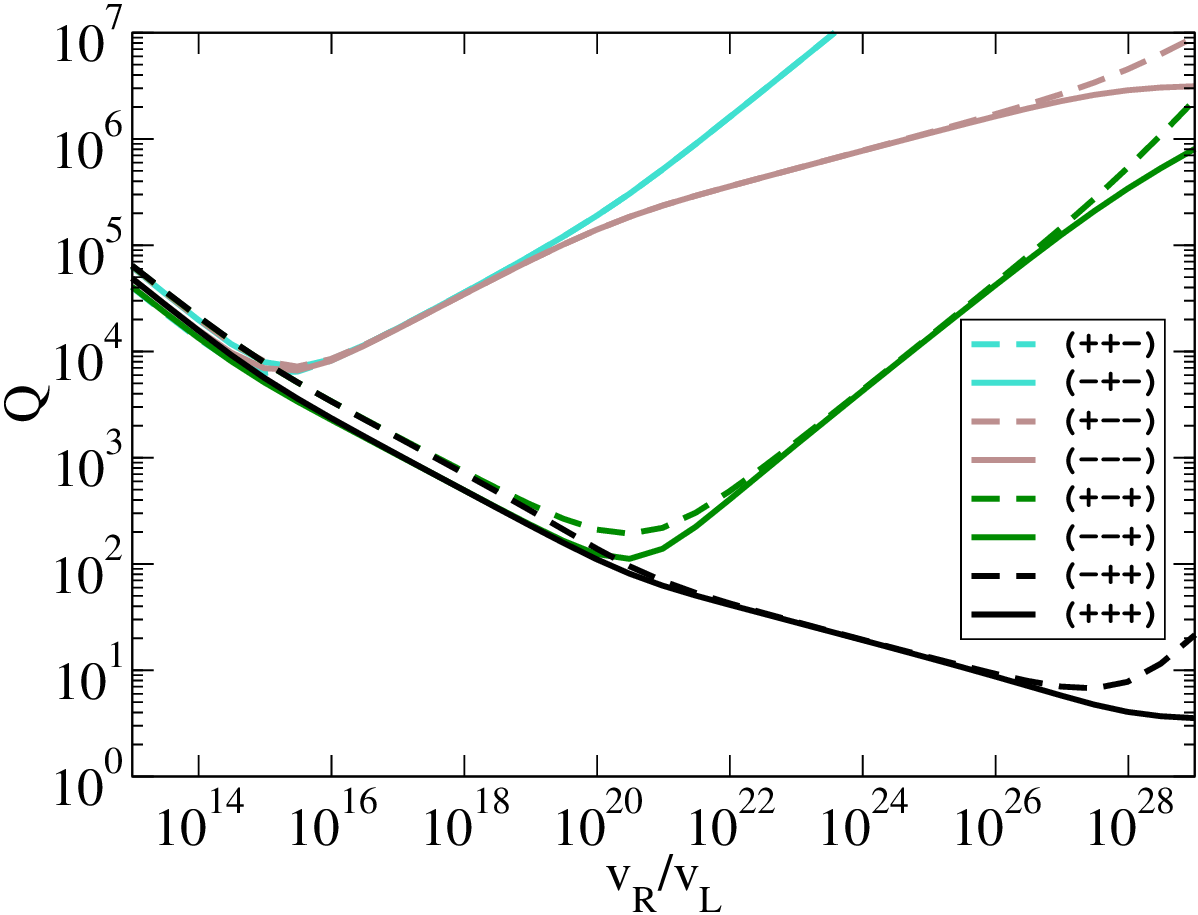}
\caption{%
\label{stab_1}
The stability measure $Q$ as a function of $v_R/v_L$ for $m_0=0.001$ 
eV. The left (right) panel corresponds to the normal (inverted) neutrino 
mass hierarchy.}
}
\FIGURE[t]{
\includegraphics[width=0.49\textwidth,clip]{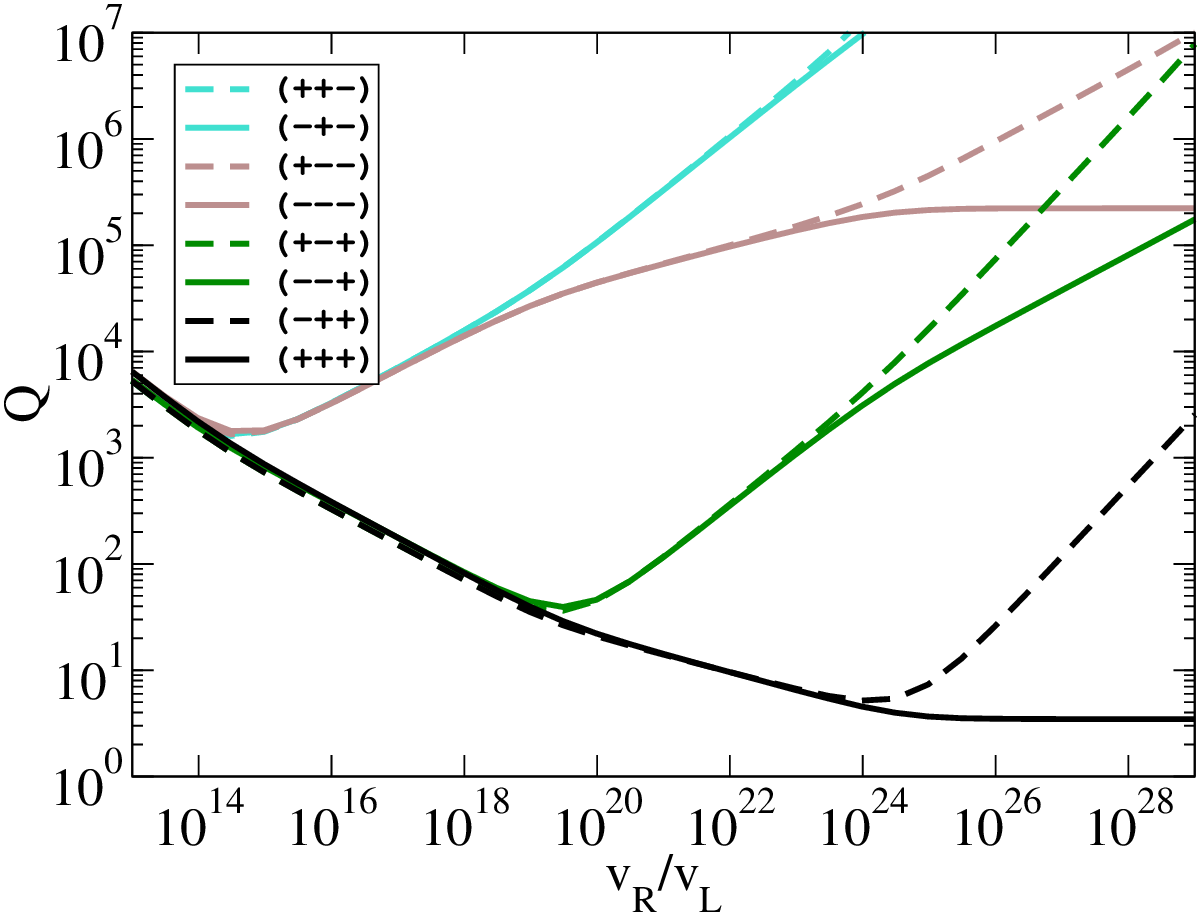}
\includegraphics[width=0.49\textwidth,clip]{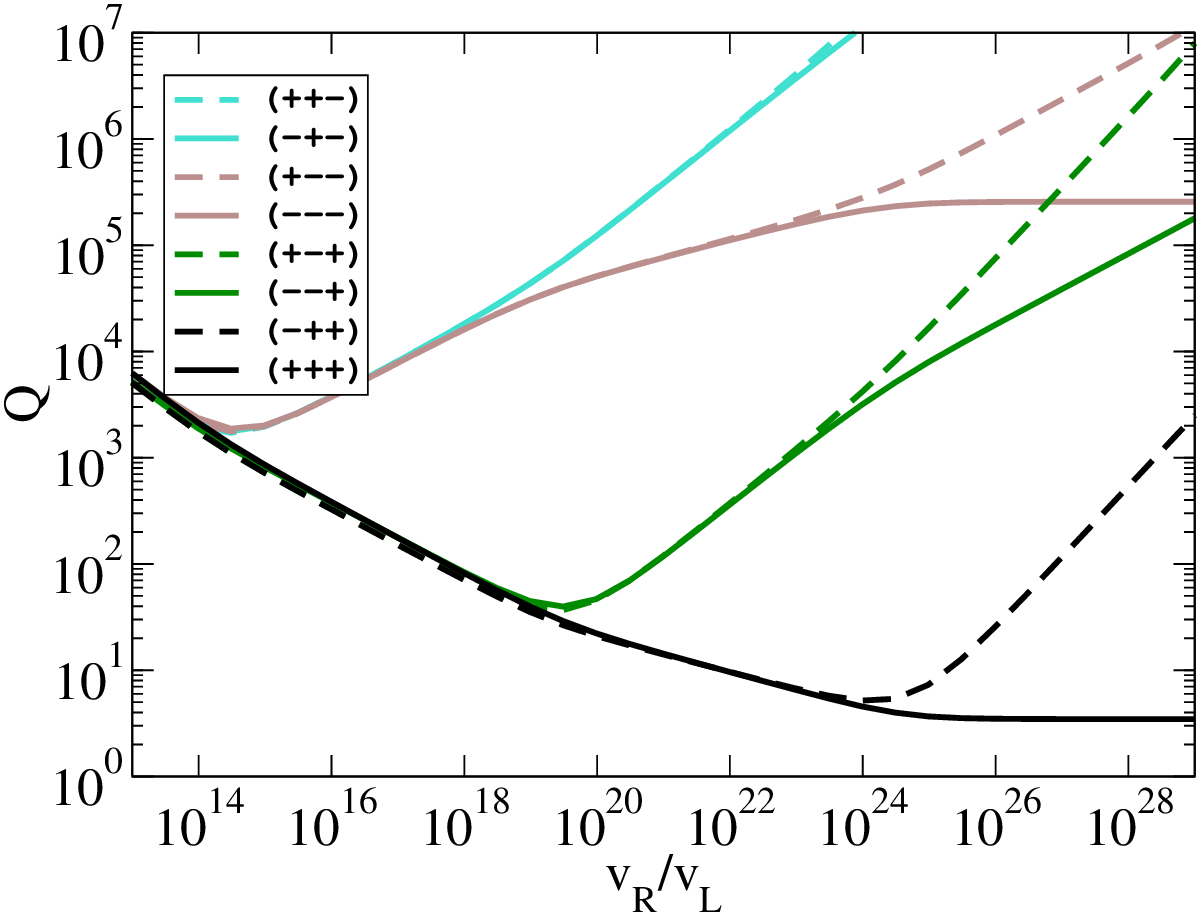}
\caption{%
\label{stab_2}
Same as in fig.~\ref{stab_1}, but for $m_0=0.1$~eV.}
}
If one allows for a tuning at a percent level, $Q \lesssim 10^3$,
then the stability analysis favors the two solutions '$\pm++$' with
$v_R/v_L \gtrsim 10^{18}$ and the two solutions '$\pm-+$' with
$v_R/v_L \simeq 10^{20}$.

It should be noted that the qualitative behavior of the stability
measure $Q$ depends mostly on the eigenvalues of the Yukawa coupling
matrix $y$ and the neutrino mass scale $m_0$. On the other hand, the
mixing in $y$ and additional CP-violating Majorana phases influence
the results only marginally.

\section{Leptogenesis}
\label{sec_lep}

In this section, we present our analysis of leptogenesis and its
implications for the discrimination among the eight allowed solutions
for $g$. Our analysis is based on the results of
refs.~\cite{Hambye:2003ka,Antusch:2004xy}.

Assuming that the lightest of the right-handed neutrinos is separated
from the other two as well as from the Higgs triplets by a large mass
gap, the baryon asymmetry arising from leptogenesis can be written as
\be
\eta_B \equiv \frac{n_B}{n_\gamma} = \eta \, \epsilon_{N_1}.
\label{eta_fac}
\ee
The observed value of the baryon asymmetry is $\eta_B = (6.1 \pm 0.2)
\times 10^{-10}$ \cite{Spergel:2003cb}. In eq.~(\ref{eta_fac}), $\eta$
is the so-called efficiency factor that takes into account the initial
density of right-handed neutrinos, the deviation from equilibrium in
their decay and washout effects, while $\epsilon_{N_1}$ denotes the
lepton asymmetry produced in the decay of the lightest right-handed
neutrino.  For the decay of the $i$th right-handed neutrino, it is
defined as
\be
\epsilon_{N_i} = \frac{\Gamma(N_i \to l\, H) - \Gamma(N_i \to \bar l\, H^*)}
	{\Gamma(N_i \to l\, H) + \Gamma(N_i \to \bar l\, H^*)}\,.
\ee
%

If the two lightest right-handed neutrinos have similar masses,
eq.~(\ref{eta_fac}) is generalized to
\be
\eta_B = \eta_1 \, \epsilon_{N_1} + \eta_2 \, \epsilon_{N_2}\,.
\ee
The coefficients $\eta_i$ mostly depend on the effective neutrino
masses, defined as
\be
\tilde m_i = \frac{v^2 \, (\hat y^\dagger \hat y)_{ii}}{2m_{N_i}}\,.
\ee
Here and below, the hat indicates that the matrices are evaluated in
the basis where the triplet Yukawa coupling matrix $g$ is diagonal
with real and positive eigenvalues. In the case of quasi-degenerate
right-handed neutrinos, $m_{N_1}\simeq m_{N_2}$, and nearly coinciding
effective masses $\tilde m_1$ and $\tilde m_2$, an order-of-magnitude
estimate of the washout coefficients gives~\cite{Pilaftsis:2005rv}
\be
\eta_i\simeq \frac{1}{200}\lp \frac{10^{-3} \, {\rm eV}}{\tilde m_i} \rp.
\ee
However, deviations from the condition $\tilde m_1 \simeq \tilde m_2$
can lead to large corrections to this estimate. In particular, a large
effective mass $\tilde m_2$ reduces the coefficient $\eta_1$ close to
the mass degeneracy point, as is shown in fig.~\ref{etaOeps}. The
results in ref.~\cite{Pilaftsis:2005rv} have been obtained for rather
light and quasi-degenerate right-handed neutrinos, $m_{N_1}\simeq
m_{N_2} \sim 1$~TeV. For hierarchical right-handed neutrino masses
$m_{N_1} \ll m_{N_2}$ and the mass scale under consideration in the
present case, $m_{N_1} \sim 10^8$ GeV, one finds
\be
\eta_1 = 1.45 \times 10^{-2}\lp \frac{10^{-3} \, {\rm eV}}{\tilde m_1} \rp, 
\qquad \eta_2 \simeq 0\,,
\ee
and we will employ these values in the following. This result and
fig.~\ref{etaOeps} have been obtained by solving the Boltzmann
equations as suggested in ref.~\cite{Pilaftsis:2005rv} and assuming
thermal initial abundance of right-handed neutrinos. 
\FIGURE[t]{
\includegraphics[width=0.90 \textwidth]{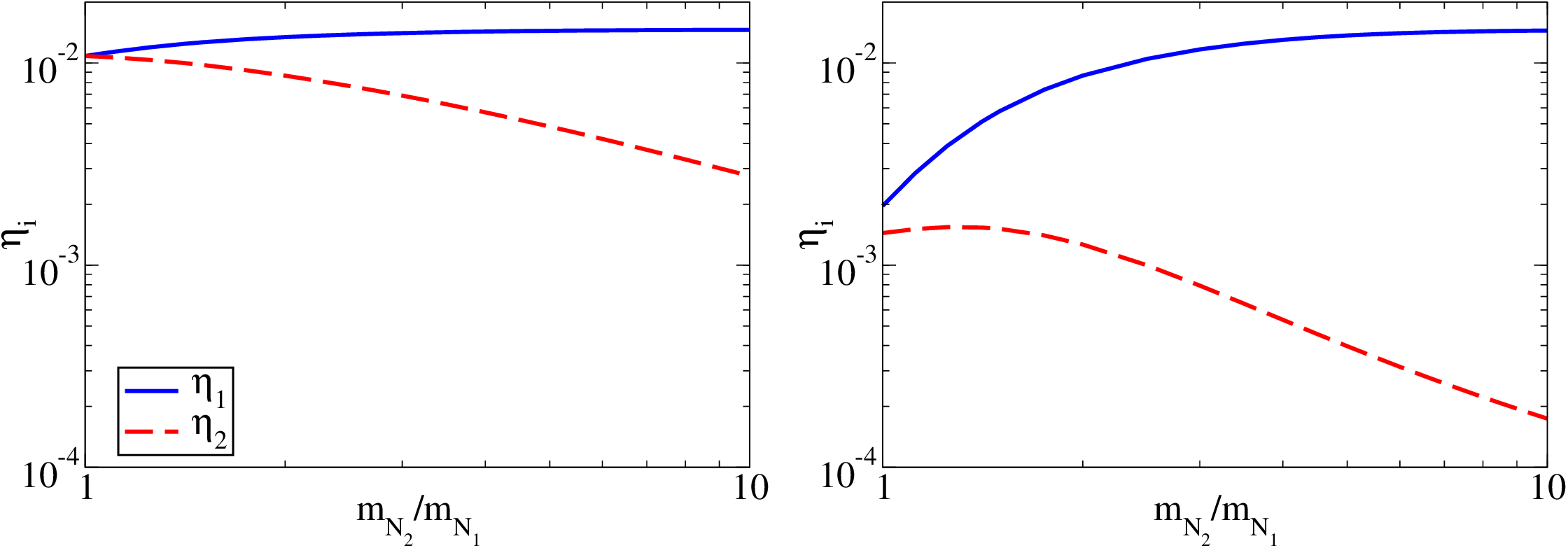}
\caption{%
\label{etaOeps}
The contributions $\eta_1$ and $\eta_2$ to the baryon-to-photon ratio
from the decays of the two lightest right-handed neutrinos versus the
ratio of their masses $m_{N_2}/m_{N_1}$. Left panel: $\tilde m_1 =
\tilde m_2 = 10^{-3}$~eV, right panel: $\tilde m_1 = 10^{-3}$~eV,
$\tilde m_2 = 10^{-2}$~eV.}  }

With the washout factors $\eta_i$ at hand, the determination of the
baryon asymmetry requires only the knowledge of the CP-violating decay
asymmetries of the right-handed neutrinos $\epsilon_{N_i}$. In the
case when the low-energy limit of the theory is the Standard Model,
$\epsilon_{N_1}$ is given by~\cite{Antusch:2004xy}
\bea
\epsilon_{N_1} &=& \epsilon^I_{N_1} + \epsilon^{II}_{N_1}\,,
\label{first}\\ 
\epsilon^I_{N_1} &=& \frac{1}{8 \pi} 
\sum_{j\not= 1} \frac{ \imag[(\hat y^\dagger \hat y)^2_{1j}]}
{ (\hat y^\dagger \hat y)_{11}} 
\sqrt{x_j} \left( \frac{2-x_j}{1-x_j} - (1+x_j)\ln{\frac{x_j+1}{x_j}} \right)\,, \\
\epsilon^{II}_{N_1} &=& \frac{3}{8 \pi} \hat g_{11} \mu \frac{\imag[(\hat 
y^\dagger \, \hat g \hat y^*)_{11}]}{(\hat y^\dagger \hat y)_{11}}
\, z \left( 1-z\,\ln{\frac{z+1}z}\,\right),
\label{lep_exact}
\eea
and analogous formulas hold for $\epsilon_{N_2}$. Here
$z=m^2_\Delta/m^2_{N_1}$, and $x_j$ is defined as the ratio of the
squared right-handed neutrino masses:
\be
x_{j} = \frac{\hat g^2_{jj}}{\hat g^2_{11}}.
\ee
In the following, we discuss only the limit of a very heavy $SU(2)_L$
Higgs triplet, $z\to \infty$, so that
\be
\epsilon^{II}_{N_1} \to \frac{3}{16 \pi} \hat g_{11} \mu
\frac{\imag[(\hat y^\dagger \, \hat g \hat y^*)_{11}]}{(\hat y^\dagger
\hat y)_{11}}.
\ee
In the limit of a strong hierarchy in the right-handed
sector, $x_j \gg 1$, the first contribution in eq.~(\ref{first}) can
be rewritten as
\be
\epsilon^I_{N_1} \to -\frac{1}{8 \pi} 
\sum_{j\not= 1}
\frac{ \imag[(\hat y^\dagger \hat y)^2_{1j}]}
{ (\hat y^\dagger \hat y)_{11}} 
\frac{3}{2\sqrt{x_j}}
= -\frac{3}{16 \pi} \hat g_{11} 
\frac{\imag[(\hat y^\dagger \, \hat y \hat g^{-1} \hat y^T \hat y^*)_{11}]}
{(\hat y^\dagger \hat y)_{11}},
\ee
so that
\be
\epsilon_{N_1} = \epsilon^I_{N_1} + \epsilon^{II}_{N_1} \to \frac{3}{16 
\pi} \hat g_{11} \mu \frac{\imag[(\hat y^\dagger \, \hat m_\nu \hat y^*)_{11}]}
{(\hat y^\dagger \hat y)_{11}}\,.
\label{lep_approx}
\ee
However, even in this limit, this approximation can lead to large
deviations from the exact result of
eqs.~(\ref{first})-(\ref{lep_exact}). Consider e.g.~the regime of
small $\mu$, where type I and type II seesaw contributions almost
cancel each other in the expression for the light neutrino mass
matrix. In this case, even a small correction to the coefficient of
the asymmetry $\epsilon^I_{N_1}$ leads to an incomplete cancellation
and to large errors in the approximation of
eq.~(\ref{lep_approx}). This effect is also partially present at
intermediate values of $\mu$. In addition, close to the mass
degeneracy ($x_j \simeq 1$), a resonant feature is expected in
$\epsilon^I_{N_1}$, which can lead to successful leptogenesis even at
a TeV scale~\cite{Pilaftsis:2005rv}.
\FIGURE[t]{
\includegraphics[width=0.99 \textwidth]{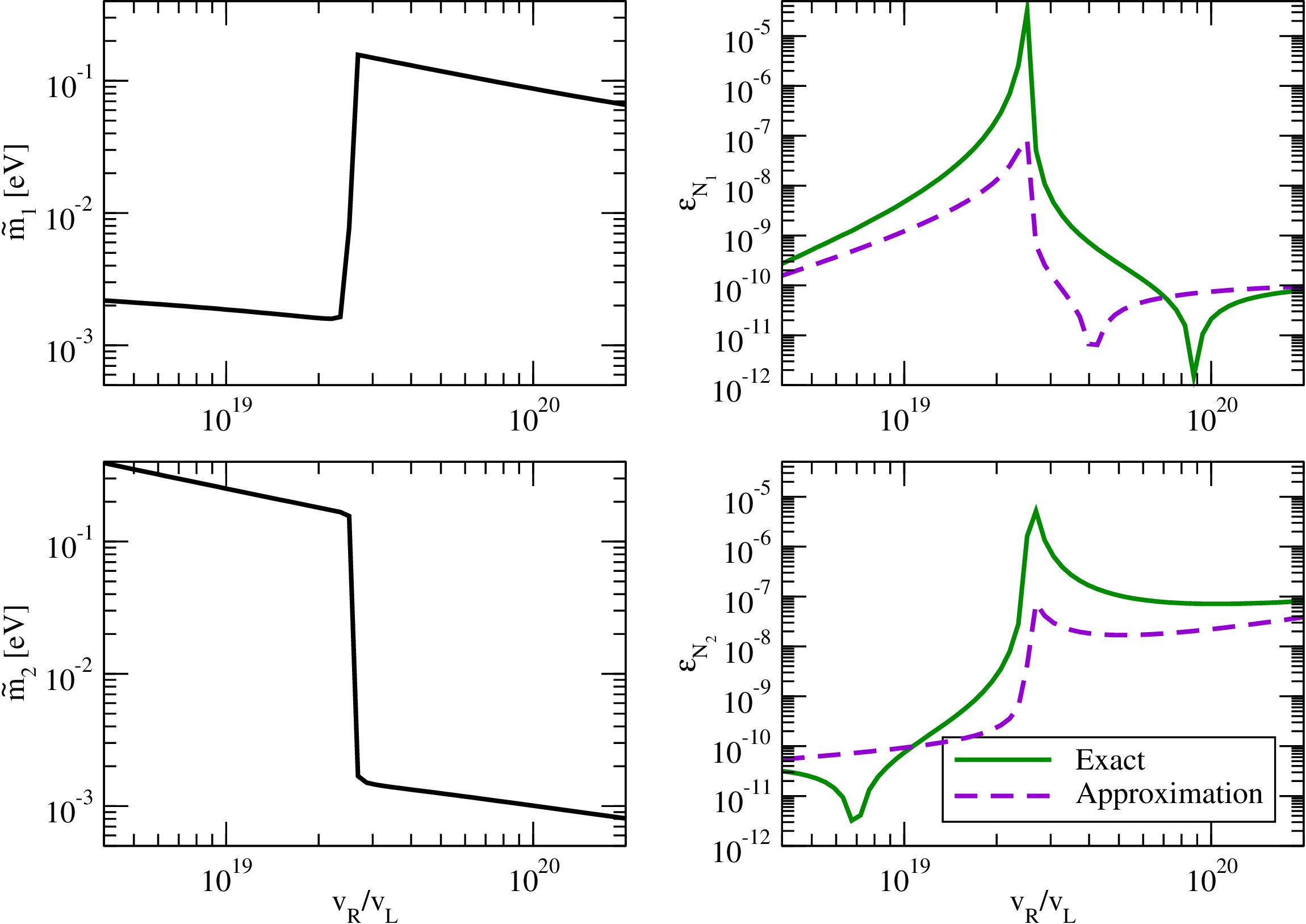}
\caption{%
\label{lep1}
The upper (lower) panels show the effective neutrino mass $\tilde m_1/
\textrm{eV}$ ($\tilde m_2/\textrm{eV}$) and the asymmetry
$\epsilon_{N_1}$ ($\epsilon_{N_2}$) as functions of $v_R/v_L$ for the
solution '$+-+$'. The dashed curves in the right panels correspond to
the approximation in eq.~(\ref{lep_approx}), while the solid curves
represent the exact result. The step-like behavior of $\tilde{m}_1$
and $\tilde{m}_2$ is due to the level crossing. Inverted mass 
hierarchy, $m_0=0.001$~eV.}
}
This is demonstrated in fig.~\ref{lep1}, where the asymmetries
$\epsilon_{N_1}$ and $\epsilon_{N_2}$ produced in the decays of the
two lightest right-handed neutrinos and the corresponding effective
mass parameters $\tilde m_1$ and $\tilde m_2$ are plotted. The results
show sizable deviations from the approximation (\ref{lep_approx}),
even outside the resonant enhancement region. The corresponding
baryon-to-photon ratio is shown in fig.~\ref{eta1}. In addition, this
figure shows the baryon-to-photon ratio in the case of non-vanishing
$\theta_{13}$ and the Dirac-type leptonic CP-violating phase
$\delta_{\rm CP}=30^\circ$. The resonant behavior is less distinct for
larger values of $\theta_{13}$, which can be traced back to the fact
that the two lightest right-handed neutrinos never become exactly
degenerate in mass in this case. On the other hand, the Dirac-type
phase constitutes an additional source of CP violation in the case of
non-vanishing $\theta_{13}$, leading to an enhancement of
$\epsilon_{N_1}$ below the mass degeneracy point for smaller values of
$\theta_{13}$, and thus, widening the $v_R/v_L$ region where 
successful leptogenesis is possible (see the dashed curve in
fig.~\ref{eta1}).

\FIGURE[t]{
\includegraphics[width=0.6 \textwidth]{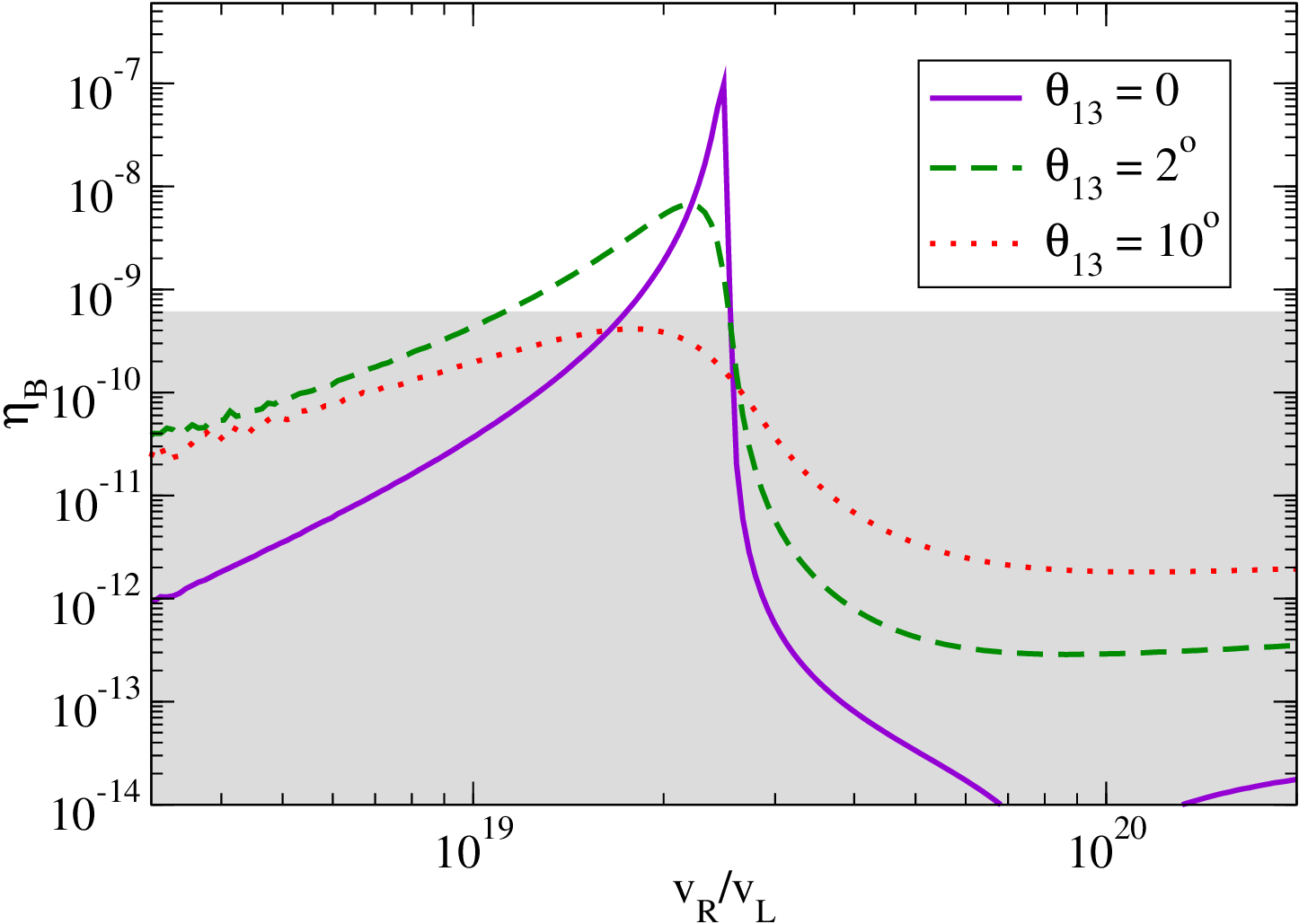}
\caption{%
\label{eta1}
The baryon-to-photon ratio $\eta_B$ from the decay of the lightest
right-handed neutrino for the solution '$+ - +$'. The same parameters
as in fig.~\ref{lep1}, except that the dashed and dotted curves
correspond to nonzero $\theta_{13}$ and $\delta_{\rm
CP}=30^\circ$. The shaded area corresponds to values of $\eta_B$ below
the observed value.}  }
Thus, we find that viable leptogenesis is possible in this scenario if
the ratio of the VEVs is close to $v_R/v_L \simeq (1 \div 2) \times
10^{19}$. Note that leptogenesis in the case of the left-right
symmetric seesaw mechanism was previously considered in a similar
framework in ref.~\cite{Hosteins:2006ja}. For the specific choice of
the parameters made there, the washout processes were found to be too
strong to allow successful leptogenesis. However, for our choice of
the parameters with the inverted mass hierarchy in the light neutrino
sector, the drop in the effective mass $\tilde{m}_1$ below the level
crossing point of the two lightest right-handed neutrinos resolves
this issue. We notice that the use of the exact formulas (4.7-4.9)
rather than the approximation (4.13) is essential in this region.

It should be also
noted that a similar effect of incomplete cancellation can appear if
the mass of the Higgs triplet is of the same order as the mass of the
lightest right-handed neutrino. In this case, the asymmetry
$\epsilon^{II}_{N_1}$ is modified and the cancellation between type I
and type II contributions is incomplete as well, which in the small
and intermediate $\mu$ regimes can enhance the produced lepton
asymmetry by several orders of magnitude compared to the approximation
in eq.~(\ref{lep_approx}).

With the parameters of fig.~\ref{eta1}, the lightest right-handed
neutrino has a mass of order $m_{N_1} \simeq 5\times 10^9$~GeV, as can
be seen in fig.~\ref{ex_3}.  Since thermal leptogenesis requires a
reheating temperature $T \gtrsim M_{N_1}$, this can potentially lead
to a tension with bounds coming from gravitino cosmology in
supersymmetric theories, namely $T \lesssim (10^7 \div
10^{10})$~GeV~\cite{Kawasaki:2004qu}. Thus, this possibility imposes
constraints which are similar to those in the usual pure type I seesaw
scenario.

Another difference from the standard leptogenesis scenario is the
appearance of the phases contained in $P_\nu$, $P_{l}$, $P_{u}$, and
$P_{d}$ in the neutrino mass matrix $m_\nu$ and in the Dirac Yukawa
coupling matrix $y$, which up to now have been set to zero in our
discussion. Due to these phases and an interplay between type I and
type II contributions to the neutrino mass matrix, leptogenesis is
possible, in principle, even in the case of one leptonic flavor, as
will be demonstrated below. This case is quite similar to the
framework with three left-handed neutrinos and one right-handed
neutrino discussed in ref.~\cite{Gu:2006wj} (see also
ref.~\cite{Akhmedov:2006de}). In the following, we will present some
analytic results for the left-right symmetric one- and two-flavor
cases, before presenting numerical results for the three-flavor case.

In the one-flavor case, the light neutrino mass is given by
\be
m_\nu = g - \frac{y^2}{\mu g}
\ee
and the lepton asymmetry produced in the decay of the heavy
right-handed neutrino is
\be
\epsilon = \frac{3}{32 \pi} \frac{{\rm Im}[\hat y^{*2} \hat
m_\nu]}{\tilde m}\,.
\ee
Once again, the hat indicates that $y$ and $m_\nu$ are in the basis
where $g$ is real and positive. It turns out that the most interesting
regime is given by large values of $\mu$ and a relative phase of
$\pi/4$ between $m_\nu$ and $y$.  In this case, only the solution
dominated by the type II term is relevant, since the type I
contribution to $\hat{y}^{*2} \hat m_\nu$ is real and cannot generate any
CP asymmetry. Thus, we obtain
\be
g \simeq m_\nu, \quad m_N = m_\nu \mu v^2\,, \\
\ee
and
\bea
\tilde m &=& \frac{|y|^2\, v^2}{2 m_N} = \frac{|y|^2}{2\, m_\nu \mu}\,,\\
\epsilon &=& \frac{3}{16 \pi} m_\nu^2 \mu = \frac{3}{16 \pi}
\frac{m_\nu m_N}{v^2}\,,\\
\eta_B &=& 1.7 \times 10^{-6} \, {\rm eV} \, \frac{m_\nu^3 \mu^2}{|y|^2}
= 1.7 \times 10^{-6} \, {\rm eV} \,  \frac{m_\nu m_N^2}{|y|^2\, v^4}\,.
\eea
Thus, it is possible to reproduce the observed baryon asymmetry
e.g.~with the values
\be
|y|=10^{-4}\,,\qquad m_\nu=0.1\; {\rm eV}\,,\qquad
\mu = 6.0 \times 10^{-5}\; {\rm eV}^{-2}\,, 
\ee
which leads to
\be
\tilde m = 8.3 \times 10^{-4} \; {\rm eV}\,, \qquad m_N = 1.8 \times 
10^8 \; {\rm GeV}\,.
\ee

The situation, however, is more complicated in scenarios with more than
one lepton flavor. For instance, mixing could give large contributions
to $\tilde m_1$, thereby enhancing the washout. On the other hand,
it can also lead to additional sources of CP violation, which might
improve the prospects for successful leptogenesis in realistic models
with several flavors. Consider, for example, the situation when the
third right-handed neutrino is much heavier than the other two and the
mixing with the third flavor in the right-handed sector is
suppressed. A novel aspect of this effective two-flavor case is that
large mixing and resonant amplification of the lepton asymmetries due
to the level crossing of right-handed neutrinos can enhance
leptogenesis. These effects are similar to those discussed above in
the full three-flavor framework.
We will study the regime with a large hierarchy between the two
lightest right-handed neutrinos, which allows a simple analytic
approach. As a toy example, we consider the following scenario: We
assume maximal mixing in the light neutrino sector and one complex
phase in $P_{l}$, which can be moved into the Yukawa coupling matrix
$y$ by rephasing the electron neutrino field. Thus, the neutrino mass
matrix is taken to have the form
\be
\label{2dex}
m_\nu = \begin{pmatrix}
e^{2i\kappa}\, \bar m & e^{i\kappa}\, \delta m \\
e^{i \kappa}\, \delta m & \bar m \\
\end{pmatrix}
\ee
with $\delta m\ll \bar m$.  The parameters $\bar m$ and $\delta m$ can
be determined from the mass of the lightest active neutrino $m_0$ and
$\Delta m^2_{21}$:
\be
\bar m \simeq m_0, \quad \delta m \simeq \frac{\Delta m^2_{21}}{4 m_0}\,.
\ee
Numerical analysis indicates that the most interesting region in the
parameter space corresponds to the situation when the smaller
eigenvalue of $g$ is in the large $\mu$ regime, while the larger
eigenvalue is in the small $\mu$ regime, i.e.
\be
\label{mu_bounds}
\frac{4 \, y_1^2}{\bar m^2} \ll \mu \ll \frac{4 \, y_2^2}{\bar m^2}\,,
\ee
and we will assume this to hold in the present example.  In this case,
two solutions for $g$ are, to first order in $\lambda$, given by the
ansatz\footnote{The other two solutions do not lead to successful
leptogenesis.}
\be
g = U^\dagger  \, \begin{pmatrix}
\bar m & 0 \\
0 & \pm \frac{y_2}{\sqrt{\mu}} + \frac{\bar m}2\\
\end{pmatrix}\, U^*, \quad 
U = \begin{pmatrix}
e^{-i \kappa} & \lambda e^{-i(\phi+\kappa)} \\
-\lambda e^{i\phi} & 1 
\end{pmatrix} 
\ee
with 
\bea
 \lambda &=& \mp \frac{\delta m \sqrt{\mu}}{y_2}\,, \\
\sin(\phi+\kappa) &\simeq& \mp \sin(2\kappa) \frac{y_1}{\bar m 
\sqrt{\mu}}\,,  
\eea
and thus, we find
\bea
\tilde m_1 &=& \frac{y_1^2 + y_2^2 \lambda^2}{2 \bar m \mu} = 
\frac{y_1^2 + \delta m^2 \mu}{2 \bar m \mu}\,, 
\label{ex2mt} \\
\epsilon_{N_1} &=& \frac{3}{32 \pi \tilde m_1} \, 
\left[ \sin(2\phi+2\kappa) \, \bar m \, \delta m^2 \mu 
+ \sin(4 \kappa) \bar m y_1^2 \right].
\label{ex2eps}
\eea
The second term in $\epsilon_{N_1}$ essentially coincides with the
corresponding expression in the one-flavor case. Hence, in this
case, it is possible to generate a sufficient lepton asymmetry in
exactly the same way as in the one-flavor case as long as the
contribution from mixing to $\tilde m_1$ does not lead to a strong
washout. The latter condition reads
\be
\frac{\delta m^2}{ 2\bar m} \simeq  \frac{(\Delta m^2_{21})^2}{ 32 m_0^3} 
\lesssim 10^{-3}~ {\rm eV}\,,
\ee
which is easily satisfied if $ m_0 > 10^{-3}$~eV.
It is interesting to note that for $\kappa=\pi/8$ and quasi-degenerate
neutrino masses, the obtained asymmetry $\epsilon_{N_1}$ saturates the
upper limit obtained in ref.~\cite{Antusch:2004xy}.

But even in the case $\kappa \simeq \pi/4$, when the second term in
the expression for $\epsilon_{N_1}$ in eq.~(\ref{ex2eps}) is
suppressed, the first term can lead to viable leptogenesis. The
corresponding contribution to $\eta_B$ takes its largest value when
$\delta m^2 = y_1^2/\mu$, so that eqs.~(\ref{ex2mt}) and
(\ref{ex2eps}) become
\bea
\tilde m_1 &=& \frac{y_1^2}{\bar m \mu}, \\ 
\epsilon_{N_1} &=& \frac{3}{16 \pi} y_1 \bar m \sqrt{\mu}.
\eea
In this case, $\eta_B$ is smaller than it is in the one-flavor case
only by a factor
\be
\frac{y_1 }{ 2\bar m\sqrt{\mu}} = \frac12 \sqrt{\frac{\tilde 
m_1}{\bar m}} \simeq 0.1. 
\ee
It should be noted that the baryon asymmetry increases with the
parameter $\mu$, so that, depending on the Yukawa couplings,
saturation of the upper limit on $\mu$ in eq.~(\ref{mu_bounds}) might
be necessary, which can lead to deviations from our analytic results.

Thus, in the two-flavor case, two different sources of leptogenesis
exist: The first source is similar to that in the one-flavor case,
which is related to the type II seesaw term and is sensitive to the
high-energy CP-violating phases, while the second source results from
mixing effects and has no analogue in the one-generation case.

In the three-flavor framework, sources of both types are, in general,
present as well, but mixing with the third flavor can further increase
$\tilde m_1$.  Figure~\ref{eta2} shows the baryon-to-photon ratio
$\eta_B$ when an additional phase is attributed to the electron
neutrino, as in the two-flavor example of eq.~(\ref{2dex}). We choose
the phase $\kappa=\pi/4$ ($\kappa=\pi/8$), so that the source similar
to the first (second) term in eq.~(\ref{ex2eps}) gives the largest
contribution to the baryon asymmetry. Our numerical results indicate that, 
similarly to the two-flavor case, the upper bound on the decay 
asymmetry found in ref.~\cite{Antusch:2004xy} can be
saturated.
\FIGURE[t]{
\includegraphics[width=0.6 \textwidth]{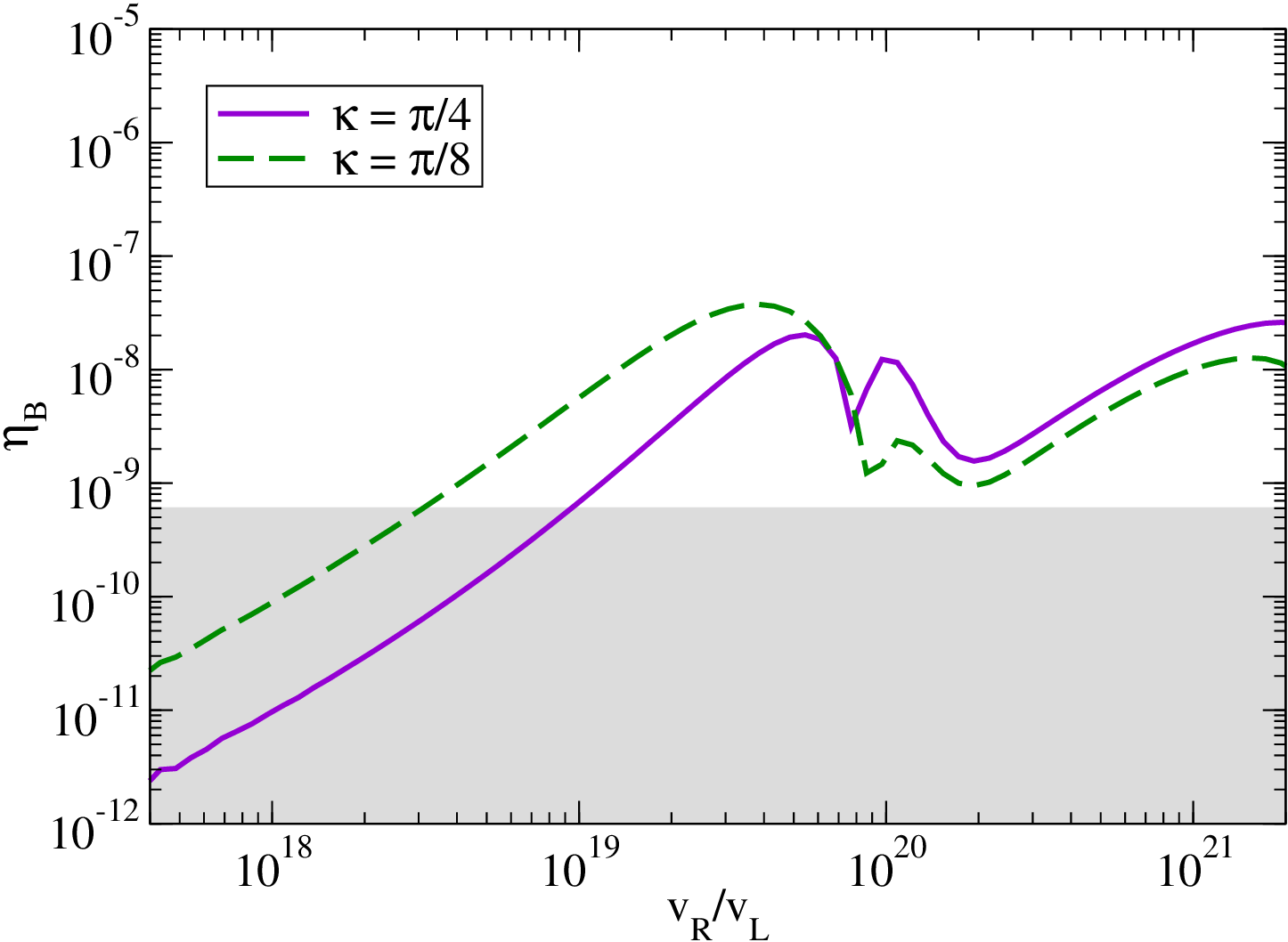}
\caption{ 
\label{eta2} 
The baryon-to-photon ratio $\eta_B$ with an additional complex phase
$\pi/8$ or $\pi/4$ attributed to the electron neutrino for the
solution '$--+$'. The shaded area corresponds to values of $\eta_B$
below the observed value. Inverted mass hierarchy, $m_0 = 0.1$~eV.}}
The mass of the lightest right-handed neutrino that is required to
reproduce the observed baryon asymmetry is $m_{N_1}\gtrsim 1.4 \times
10^9$~GeV ($m_{N_1}\gtrsim 2.5\times 10^8$~GeV). These bounds can be
relaxed by choosing Yukawa couplings different from those of the
up-type quarks. With an appropriate choice, the results for the four
solutions of the type '$\pm\pm+$' agree with the analytic predictions
of the two-flavor analysis presented in this section. Notice that the
results in the two-flavor case in eqs.~(\ref{ex2mt}) and
(\ref{ex2eps}) do not depend on $y_2$ as long as the constraint
(\ref{mu_bounds}) is fulfilled. Likewise, we observe in the numerical
analysis of the three-flavor case that in this limit leptogenesis is
not very sensitive to the two largest eigenvalues $y_2$ and
$y_3$. This is, however, a consequence of the fact that the mixing in 
the 1-3 sector of the Dirac-type Yukawa coupling $y$ is small in our
framework according to eq.~(\ref{ckm_2}). If this mixing is sizable, 
$\theta^q_{13} \gtrsim 5^\circ$, and depending on the other parameters 
determining the Yukawa coupling $y$ and the
neutrino mixing matrix $U_{\rm PMNS}$, leptogenesis might be
suppressed, mainly due to a large contribution to the effective mass
parameter $\tilde m_1$ from the eigenvalue $y_3$ and the 
resulting increased washout.

Thus, we conclude that successful leptogenesis is possible for four
out of the eight solutions provided that the value of the
electron-type Majorana phase is in an appropriate range. For the other
four solutions, leptogenesis is not viable, as was first pointed out
in ref.~\cite{Hosteins:2006ja}. The reason for this is that, as long
as the Dirac Yukawa coupling matrix is chosen to coincide with that of
the up-type quarks, the mass of the lightest right-handed neutrino
never exceeds $10^6$~GeV and no level crossings occur. We note that in
the left-right symmetric case with type I+II seesaw mechanism the
bounds on the mass of the lightest right-handed neutrino can be
slightly relaxed compared to those in the pure type I case which, for
right-handed neutrinos with thermal initial abundance and hierarchical
masses, requires $m_{N_1}\gtrsim 5\times
10^8$~GeV~\cite{Davidson:2002qv,Buchmuller:2002rq,Giudice:2003jh}.

\section{Summary and conclusions\label{sec_summary}}

\TABLE{
\begin{tabular}[b]{|c||c|c|c|}
\hline
 & $\pm++$ & $\pm-+$ & $\pm\pm-$ \\
\hline
\hline
Stability & $v_R/v_L > 10^{18}$ & $v_R/v_L \simeq 10^{20}$ &  disfavored \\
\hline
\quad Leptogenesis \quad & $v_R/v_L > 10^{18}$ & $v_R/v_L > 10^{18}$ &
excluded \\
\hline
Gravitinos & \quad $v_R/v_L < 10^{21}$ \quad & \quad unconstrained
\quad & \quad unconstrained \quad \\
\hline
\end{tabular}
\caption{The allowed regions of the parameter $v_R/v_L$ for the eight
different types of solutions.\label{finaltab}}
}

We have analyzed the left-right symmetric type I+II seesaw mechanism
with a hierarchical Dirac mass term motivated by GUTs. It was
previously shown that a reconstruction of the mass matrix of heavy
right-handed neutrinos in this framework produces eight solutions
which result in exactly the same low-energy phenomenology. Our goal
was to discriminate among these solutions using their stability
properties and leptogenesis as additional criteria.
As a measure of the stability, we have chosen the parameter $Q$ which
quantifies the degree of fine-tuning necessary to obtain a given mass
matrix of light neutrinos and was defined in
eq.~(\ref{Q_measure}). For three lepton generations, no fine-tuning
corresponds to $Q\sim 10$. We have selected the value $Q=10^3$, which
corresponds to a fine-tuning at the percent level, as a maximal
allowed value. The leptogenesis criterion we used was the ability of
a given solution to reproduce the observed baryon asymmetry of the
Universe.

Our results complement the results of the leptogenesis analysis
performed in ref.~\cite{Hosteins:2006ja} in the following aspects. In
the case without additional Majorana phases, we obtain, in accordance
with ref.~\cite{Hosteins:2006ja}, that a sizable decay asymmetry
$\epsilon_{N_1}$ is possible close to the mass degeneracy of the two
lightest right-handed neutrinos. However, while for the specific
parameters used in ref.~\cite{Hosteins:2006ja} the washout is too
large to allow viable leptogenesis, we find that assuming the inverted
mass hierarchy for the light neutrinos resolves the problem, as shown
in fig.~\ref{lep1}. Similarly, in the cases with additional
CP-violating Majorana phases we found that for certain solutions the
choice of the parameters made in ref.~\cite{Hosteins:2006ja} leads
either to a strong washout (solutions '$\pm-+$'), or to a violation of
the gravitino bound (solutions '$\pm++$'). In section~\ref{sec_lep},
we presented a systematic study showing that those problems can be
solved for the four solutions '$\pm\pm+$' if the value of the of
electron-type Majorana phase is in the appropriate range. In
particular, the upper bound on the decay asymmetry for the type I+II
seesaw model found in ref.~\cite{Antusch:2004xy} can be saturated for
a certain choice of the parameters. This is illustrated by the
analytic results for the two-flavor case in eqs.~(\ref{ex2mt}) and
(\ref{ex2eps}) and the numerical results for the three-flavor case in
fig.~\ref{eta2}. We would like to emphasize that if the Dirac-type Yukawa
coupling matrix $y$ is characterized by hierarchical eigenvalues and rather 
small mixing, successful leptogenesis is quite a generic feature of the 
left-right symmetric seesaw models. 

Our findings are summarized in tab.~\ref{finaltab}. One can observe
that the stability criterion disfavors the four solutions of the type
'$\pm\pm-$' and restricts the solutions of the type '$\pm-+$' to the
region of the parameter space where $v_R/v_L \simeq 10^{20}$. The
remaining two solutions of the type '$\pm++$' are stable, provided
that $v_R/v_L \gtrsim 10^{18}$. We found that successful leptogenesis
is possible for the four solution of the type '$\pm\pm+$' as long as
$v_R/v_L \gtrsim 10^{18}$. This possibility requires the existence of
additional Majorana-type phases which are absent in the pure type I
seesaw framework. Further constraints come from the potentially
dangerous overproduction of gravitinos in supersymmetric theories,  
giving rise to an upper bound on the lightest right-handed
neutrino mass. For our choice of the Yukawa couplings, $y=y_u$, only
the solutions of the type '$\pm++$' are affected by this constraint,
which leads to the requirement $v_R/v_L \lesssim 10^{21}$. For the other 
six solutions, the smallest right-handed neutrino mass is always below 
$10^{10}$~GeV, so that these solutions are not constrained by this criterion. 
In the cases when the middle eigenvalue of $y$ is chosen to be 
significantly larger than the one in our framework, $y_2\gtrsim 10^{-2}$, 
the constraint $v_R/v_L \lesssim 10^{21}$ would also apply to the two 
solutions of the form '$\pm-+$'. On the other hand, a very small middle 
eigenvalue, $y_2\lesssim 5 \times 10^{-4}$, would render leptogenesis 
impossible for these two solutions, since the decay asymmetry would be too 
small due to the small mass of the lightest right-handed neutrino.

Thus, we have shown, within the chosen framework, that the stability
and leptogenesis criteria partially lift the eight-fold degeneracy
among the solutions for the mass matrix of heavy right-handed
neutrinos in the left-right symmetric type I+II seesaw.

\acknowledgments
We thank S. Lavignac and C. Savoy for useful communications. 
This work was supported by the Wenner-Gren Foundation [E.A.], the
G\"oran Gustafsson Foundation [T.H. and T.O.], the Swedish Research
Council (Vetenskapsr{\aa}det), contract nos. 621-2001-1611 [T.K. and
T.O.] and 621-2005-3588 [T.O.], and the Royal Swedish Academy of
Sciences (KVA) [T.O.].

\bibliographystyle{JHEP}
\bibliography{references}

\end{document}